\documentclass[12pt]{article}
\usepackage{amsmath, amsthm, amsfonts}
\usepackage{graphicx,psfrag,epsf}
\usepackage{enumerate}
\usepackage{natbib}
\usepackage{url} 
\usepackage{booktabs}
\usepackage{multirow}
\usepackage{verbatim}
\usepackage{setspace}
\usepackage{authblk}
\RequirePackage[colorlinks,citecolor=blue,urlcolor=blue]{hyperref}

\newcommand{\blind}{1}

\newcommand{\bmu}{\mbox{\boldmath $\mu$}}
\newcommand{\bbeta}{\mbox{\boldmath $\beta$}}

\newcommand{\bSigma}{\mbox{\boldmath $\Sigma$}}

\newcommand{\bgamma}{\mbox{\boldmath $\gamma$}}

\newcommand{\btau}{\mbox{\boldmath $\tau$}}

\begin{document}

\def\spacingset#1{\renewcommand{\baselinestretch}
{#1}\small\normalsize} \spacingset{1}

\if1\blind
{
  \title{\bf Bayesian Bivariate Subgroup Analysis for Risk-Benefit Evaluation}
  
\author[1]{Nicholas C. Henderson}
\author[1,2]{Ravi Varadhan}
\affil[1]{ {\small Sidney Kimmel Comprehensive Cancer Center, Johns Hopkins University } }
\affil[2]{ {\small Department of Biostatistics, Bloomberg School of Public Health, Johns Hopkins University} }

\date{\vspace{-6ex}}
  \maketitle
} \fi

\if0\blind
{
  \bigskip
  \bigskip
  \bigskip
  \begin{center}
    {\LARGE\bf Doubly Compromised Estimation ...}
\end{center}
  \medskip
} \fi

\bigskip

\begin{abstract}
Subgroup analysis is a frequently used tool for evaluating heterogeneity of treatment effect and heterogeneity in treatment harm across observed baseline patient characteristics. While treatment efficacy and adverse event measures are often reported separately for each subgroup, analyzing their within-subgroup joint distribution is critical for better informed patient decision-making. In this paper, we describe Bayesian models for performing a subgroup analysis to compare the joint occurrence of a primary endpoint and an adverse event between two treatment arms. Our approaches emphasize estimation of heterogeneity in this joint distribution across subgroups, and our approaches directly accommodate subgroups with small numbers of observed primary and adverse event combinations. In addition, we describe several ways in which our models may be used to generate interpretable summary measures of benefit-risk tradeoffs for each subgroup.  The methods described here are illustrated throughout using a large cardiovascular trial ($N = 9,361$) investigating the efficacy of an intervention for reducing systolic blood pressure to a lower-than-usual target.
\end{abstract}

\noindent%
\vfill

\spacingset{1.45}

\newpage

\section{Introduction} 
Both treatment effectiveness and treatment safety often vary according to patient sub-populations that are defined by observable baseline patient characteristics. Although the importance of evaluating the consistency of treatment effectiveness is often recognized when conducting subgroup analyses, heterogeneity in treatment safety is not examined as frequently. 
Moreover, when heterogeneity in treatment safety is addressed, such subgroup analyses are typically performed separately from the heterogeneity of treatment effect (HTE) analysis.
While useful for addressing question of HTE and heterogeneity in adverse effects, such separate analyses ignore potentially important relationships between primary outcomes and safety outcomes. Dependencies between these two outcomes can substantially alter the risk-benefit considerations when compared with just analyzing their marginal distributions. In order to improve the relevance of subgroups analysis for assessing variation in patients' risk-benefit profiles, it is critical to focus on examining differences in joint patient outcomes (\cite{Evans:2016}) within key patient sub-populations. 

Over the course of a clinical trial, patients experience a collection of outcomes of which some may be related to treatment benefit while others may be related to an adverse effect of treatment.
From a patient-level perspective, a risk-benefit assessment involves examining the potential set of outcomes that would occur when taking a proposed treatment versus the potential set of outcomes that would occur under a control treatment. A treatment may be considered superior to the control if the likely set of outcomes under treatment is better than the likely set of outcomes under the control. In this article, we focus on the frequently occurring case where the recorded set of patient outcomes includes a time to some primary event and a binary indicator of whether or not a treatment-related adverse event occurred at some point during the period of patient follow-up. Similar ways of addressing both efficacy and safety have been proposed, for example, in the analysis of dose-finding studies (\cite{berry:2012}), but to our knowledge, little work has been done in the context of bivariate subgroup analysis.

Methods based on Bayesian hierarchical models have a number of advantages when conducting subgroup analyses (\cite{jones:2011} or \cite{henderson:2016}), and many of these advantages should be more strongly felt when extending such methods to bivariate patient outcomes. 
It is well-recognized that subgroups defined by multivariate patient characteristics often have many subgroups with small within-subgroup samples sizes. This leads to very noisy estimates of subgroup-specific quantities - a phenomenon which is likely to be further exacerbated when examining the joint distribution of patient outcomes. A key advantage of a Bayesian approach coupled with hierarchical modeling is that subgroup-specific parameter estimates are partially driven by data in that particular subgroup and partially driven by patient outcomes in all the other subgroups. Allowing such ``borrowing of information'' across subgroups leads to shrunken estimates of subgroup parameters which frequently improves both the accuracy and stability of estimation (\cite{efron:1973}). In addition to simply generating more stable shrinkage estimates, Bayesian hierarchical models allow one to specify how subgroup parameters are related to one another. In this article, we focus on two approaches for modeling the relationships between the subgroup parameters (a saturated and an additive model), but many other alternatives could easily be incorporated into our framework for bivariate subgroup analysis.

In this article, our main goal is to harness the advantages of Bayesian modeling to describe the joint distribution of a primary and an adverse event across patient subgroups. To this end, we outline a within-subgroup parametric model which assumes an exponential distribution for both adverse event-free and adverse-event-occurring survival. While such a parametric assumption may not be strictly correct, our assumption is that survival follows an exponential distribution within each subgroup rather than the entire population represented by the trial. We argue that this is a sensible choice given the potentially small number of patients within any particular subgroup. Moreover, our approach for direct modeling of patients' joint distributions provides great flexibility to analyze patients' risk-benefit tradeoffs from a variety of perspectives. For example, we demonstrate how our model may be used to assess heterogeneity with respect to either a composite outcome which weights adverse event-free and adverse event-occurring survival differently, and how our model may be used to evaluate heterogeneity with respect to a treatment effect based on the probability of achieving an improved outcome.

This paper has the following organization. In Section 2, we begin by describing our motivating example - the SPRINT trial, and we describe both the subgroups to be analyzed and the primary and adverse events measured in this large clinical trial. In Section 3, we outline our parametric model for the joint distribution of the time-to-a-primary event and a binary adverse event, and we discuss its use in the context of bivariate subgroup analysis. Section 4 describes a general approach for specifying a prior distribution for the model parameters of Section 3, and we describe three particular ways of modeling the distribution of these parameters. In Section 5, we detail several interpretable measures of variation in patients' risk-benefit profiles. Here, we describe how our Bayesian framework may be used to make inferences about such measures. Patient outcomes from the SPRINT trial are used to illustrate the use of these measures. Section 6 discusses the use of model diagnostics and model comparison, and Section 7 concludes with a brief discussion.

\section{The SPRINT Trial}
The Systolic Blood Pressure Intervention (SPRINT) trial (\cite{nejm:2015}) was a large trial investigating the use of a lower-than-usual systolic blood pressure target among individuals deemed to be at increased cardiovascular risk. Specifically, the trial was designed to compare an intensive treatment (a systolic blood-pressure target of less than 120 mm Hg) versus the standard treatment (a systolic blood-pressure target of less than 140 mm Hg). In total, $9361$ individuals were enrolled in the trial, and each trial participant was classified as being at increased cardiovascular risk but without diabetes and had a systolic blood pressure greater than $130$ mm Hg.   
Of the $9361$ trial participants, $4678$ were randomized to the intensive treatment arm while $4683$ were randomized to the standard treatment arm. 

In the SPRINT trial, patient outcomes were recorded as a composite outcome where a primary event (PE) was said to occur if any one of the following five events took place: death from cardiovascular causes, stroke, myocardial infarction, acute coronary syndrome not resulting in myocardial infarction, or acute decompensated heart failure. At the conclusion of the trial (median follow up time of $3.26$ years), a total of $562$ PEs had occurred with $243$ PEs occurring in the intensive treatment arm and $319$ PEs occurring in the standard treatment arm. Based on these differences in the distribution of the PEs across treatment arms, the intervention was determined to have an overall benefit (an estimated log-hazard of $0.75$, confidence interval: $[0.64, 0.89]$).
In addition to the primary outcome, participants were also monitored for the occurrence of serious adverse events (SAEs). The most common SAEs included acute kidney injury or acute renal failure, injurious fall, electrolyte abnormality, syncope, and hypotension. Each of the observed SAEs 
were classified according to whether or not they were possibly or definitely
related to the intervention. Of the observed treatment-related SAEs, $220$ individuals in the intensive treatment group experienced a treatment-related SAE while only $118$ participants in the standard treatment group experienced a treatment-related SAE. 

While the marginal occurrence of SAEs in each treatment arm suggests that the standard treatment carries a lower risk of SAEs, the joint occurrence of primary events and SAEs may be more informative from a patient-centered perspective. That is, the risk-benefit tradeoffs involved in choosing a treatment strategy are more apparent when the likelihood of each PE-SAE combination can be examined. This can be particularly important in cases where the dependence pattern between the PE and SAE differs substantially across treatment arms.

\begin{table}[ht]
\centering
\begin{tabular}{@{}lrrrr@{}} \toprule
\multicolumn{1}{c}{} & \multicolumn{2}{c}{Standard Treatment} & \multicolumn{2}{c}{Intensive Treatment} \\ \cmidrule(r){2-3} \cmidrule(r){4-5}
 & SAE & No SAE  & SAE & No SAE \\ \midrule
PE    & 18  & 301  & 30  & 213 \\
No PE & 100 & 4264 & 190 & 4245 \\ \bottomrule
\end{tabular}
\caption{SPRINT trial: joint counts for primary events (PE) and serious adverse events (SAE) that were classified as possibly or definitely related to the intervention.}
\label{tab:sprint_joint}
\end{table}

Table \ref{tab:sprint_joint} shows the number of joint occurrences of PEs and treatment related SAEs in the SPRINT trial.
As evidenced by the counts in Table \ref{tab:sprint_joint}, most patients in both treatment arms remained free from both PEs and SAEs over the course of the trial. A notable difference between the two treatment arms is the difference in the proportion of PEs that were accompanied by an SAE. 
However, only examining the joint counts of PEs and SAEs, of course ignores the timings of the PEs. In the next section, we outline a subgroup-specific parametric model for describing the joint distribution of a time-to-event primary outcome and a binary safety outcome.

\section{Bivariate Subgroup Analysis}
\subsection{Data Structure and Summary Statistics}
We assume that $n$ patients have been enrolled in a randomized clinical study where patients are monitored to determine whether they experience either a primary event (PE) or an adverse event (AE) (or both).
For the $i^{th}$ individual in the study, $T_{i}$ denotes the time-to-failure for the PE of interest. We observe $Y_{i} = \min\{ T_{i}, C_{i} \}$ and an event indicator $\delta_{i} = I( T_{i} \leq C_{i})$ where $C_{i}$ denotes time-to-censoring and $I(\cdot)$ denotes the indicator function. In addition to observing the pair $(Y_{i},\delta_{i})$, we observe an indicator $W_{i}$ of whether or not patient $i$ experienced at least one AE during the study. That is, $W_{i} = 1$ if patient $i$ experienced at least one AE and $W_{i} = 0$ otherwise.
We let $A_{i} = 1$ denote that patient $i$ was assigned to the treatment arm, and we let $A_{i} = 0$ denote that patient $i$ was assigned to the control arm. We assume that $G$ distinct patient subgroups have been defined according to multivariate patient characteristics and that $G_{i}$ is a variable which indicates membership in one of the $G$ subgroups. For example, suppose that patient subgroups have been formed on the basis of age (young/old) and sex (male/female) combinations. In this case, we would have $G = 4$ subgroups to represent each age/sex combination, and $G_{i}$ would be a variable indicating to which of the four categories the $i^{th}$ individual belongs.

Subgroup analyses that only examine the primary outcome typically utilize a collection of summary statistics which measure treatment effectiveness in each subgroup considered. Such summary statistics are usually computed for marginally defined subgroups (i.e., subgroups defined by looking at one variable at-a-time) or subgroups which are defined according multivariate characteristics. In our analysis of bivariate patient outcomes, we require that one compute a pair of summary statistics $(D_{awg}, U_{awg})$ for each AE-treatment arm combination within each subgroup. The summary statistic $D_{awg}$ is the number of PEs that are observed to occur among the patients in subgroup $g$ $(g=1,\ldots,G)$, treatment arm $a$ $(a=0,1)$, and have AE status $w$ $(w=0,1)$.
Similarly, $U_{awg}$ is the total follow-up time for the group of patients that are in subgroup $g$, treatment arm $a$, and have AE status $w$. In addition to $(D_{awg}, U_{awg})$, we require that one compute the summary statistics $V_{ag}$ for each treatment arm-subgroup combination. The summary statistic $V_{ag}$ is the total number of AEs observed for those patients in subgroup $g$ and treatment arm $a$. More formally, $(D_{awg}, U_{awg})$ and $V_{ag}$ are defined in terms of the individual-level responses as
\begin{eqnarray}
D_{awg} &=& \sum_{i=1}^{n} \delta_{i} I(A_{i}=a)I(W_{i}=w)I(G_{i}=g) \nonumber \\
U_{awg} &=& \sum_{i=1}^{n} Y_{i}I(A_{i}=a)I(W_{i}=w)I(G_{i}=g) \nonumber \\
V_{ag} &=& \sum_{i=1}^{n} W_{i}I(A_{i} = a)I(G_{i} = g) \nonumber
\end{eqnarray}

Table \ref{tab:sprint_summaries} shows the summary statistics $(D_{awg}, U_{awg}, V_{ag})$ from the SPRINT trial using $G=8$ subgroups. These $8$ subgroups were formed by looking at all possible combinations of the following three baseline patient variables: chronic kidney disease (Yes/No), age ($\geq 75/< 75$), and sex (male/female).

\begin{table}[ht]
\centering
\begin{tabular}{@{}lllccrccrr@{}} \toprule
\multicolumn{3}{c}{Subgroup} & \multicolumn{3}{c}{Standard Treatment} & \multicolumn{3}{c}{Intensive Treatment} & \multicolumn{1}{c}{} \\ \cmidrule(r){4-6} \cmidrule(r){7-9}
CKD & Sex & Age & $(D_{01g}, U_{01g})$ & $(D_{00g}, U_{00g})$ & $V_{0g}$ & $(D_{11g}, U_{11g})$ & $(D_{10g}, U_{10g})$ & $V_{1g}$ & $N_{g}$ \\ \midrule
 No & $< 75$ & Male & (3, 92.00) & (96, 5546.11) & 29 & (10, 174.05) & (54, 5446.53) & 61 & 3528  \\ 
 Yes & $< 75$ & Male & (3, 48.41) & (28, 1364.73) & 16 & (5, 103.15) & (25, 1227.70) & 32 & 858 \\ 
 No & $\geq 75$ & Male & (1, 19.57) & (46, 1277.47) & 8 & (1, 57.67) & (26, 1265.94) & 19 & 913 \\ 
 Yes & $\geq 75$ & Male & (6, 40.00) & (47, 977.71) & 15 & (7, 88.31) & (38, 974.85) & 31 & 730 \\ 
 No & $< 75$ & Female & (0, 40.08) & (31, 2641.97) & 13 & (0, 67.47) & (25, 2705.38) & 20 & 1706 \\ 
 Yes & $< 75$ & Female & (2, 42.41)  & (12, 948.66) & 14 & (4, 50.23) & (16, 1000.35) & 16 & 617 \\ 
 No & $\geq 75$ & Female & (0, 37.84)  & (16, 835.07) & 12  & (1, 60.98)  & (18, 778.16) & 20 & 568 \\ 
 Yes & $\geq 75$ & Female & (3, 30.81)  & (25, 612.09) & 11  & (2, 66.44) & (11, 634.87) & 21 & 441 \\ 
\bottomrule
\end{tabular}
\caption{SPRINT trial: Summary statistics for the $8$ subgroups defined by the baseline variables: chronic kidney disease (CKD), age, and sex. The summary statistics $U_{awg}$ are computed using time measurements in years, but these are not provided here.}
\label{tab:sprint_summaries}
\end{table}

\subsection{Distribution of Key Summary Statistics and Joint Distribution of Primary and Adverse Events}
In our analysis, we assume the number of observed events $D_{awg}$ follows a Poisson distribution whose mean depends on the total follow-up time $U_{awg}$ and the hazard rate $\lambda_{awg}$ within the subset of patients with the $(a,w,g)$ combination of treatment arm $a$, AE status $w$, and subgroup $g$. Specifically, given $\lambda_{awg}$, we assume $D_{awg}$ is distributed as
\begin{equation}
D_{awg}|\lambda_{awg} \sim \textrm{Poisson}(\lambda_{awg}U_{awg}).
\nonumber
\end{equation}
The assumed Poisson distribution for the summary statistics $D_{awg}$ is equivalent to assuming that, at the individual level, the time-to-failure $T_{i}$ of the primary event follows an exponential distribution with rate $\lambda_{awg}$. More specifically, the likelihood function that results from assuming $D_{awg}|\lambda_{awg} \sim \textrm{Poisson}(\lambda_{awg}U_{awg})$ is the same (ignoring constant terms) as the likelihood function that results from assuming that
\begin{equation}
T_{i}|W_{i}=w, A_{i} = a, G_{i}=g \sim \textrm{Exponential}(\lambda_{awg}). \nonumber
\end{equation}

To fully define the joint distribution of the time-to-primary event $T_{i}$
and the AE indicator $W_{i}$ given treatment arm and subgroup information, we now only need to specify the conditional distribution of $W_{i}$ given $A_{i}$ and $G_{i}$. Here, we assume that $W_{i}$ is a Bernoulli random variable with a success probability $p_{ag}$ that depends on the treatment arm and subgroup. This implies the distribution of $V_{ag}$ is given by
\begin{equation}
V_{ag}|p_{ag} \sim \textrm{Binomial}(n_{ag}, p_{ag}), \nonumber 
\end{equation}
where $n_{ag}$ is the number of individuals in treatment arm $a$ and subgroup $g$.

The parameters $(\lambda_{awg}, p_{ag})$ characterize the joint distribution of the primary and adverse event within the subgroup $g$ and treatment arm $a$. To see why this is the case, note that the joint probability $P(T_{i} > t, W_{i}=w|A_{i}=a, G_{i}=g)$ may be expressed as
\begin{eqnarray}
S_{ag}(t,w) &=&
P(T_{i} > t, W_{i}=w|A_{i}=a, G_{i}=g) \nonumber \\
&=& P(T_{i} > t|W_{i}=w, A_{i}=a, G_{i}=g)P(W_{i}=w|A_{i} = a, G_{i}=g) \nonumber \\
&=& e^{-\lambda_{awg}t}p_{ag}^{w}(1 - p_{ag})^{1 - w}. \nonumber
\end{eqnarray}
Comparing $S_{1g}(t, w)$ and $S_{0g}(t, w)$ for each subgroup enables one to compare the effect of the treatment on altering the risk-benefit profiles of patients in subgroup $g$.

\section{Modeling Subgroup Effects}
\subsection{Regression Models for Subgroup Parameters}
As described above, our model for the summary measures $(D_{awg}, U_{awg}, V_{ag})$ depends on the collection of hazard rate parameters $\lambda_{awg}$ and the AE probabilities $p_{ag}$. Because the number of parameters is potentially quite large, models which induce shrinkage are needed to produce stable estimates of these quantities. We describe a general regression models for $\lambda_{awg}$ and $p_{ag}$ which allow for shrinkage and allow for one to include information regarding how the subgroups parameters are related. In Sections 4.1.1-4.1.3, we describe three specific regression models, and in Section 4.2 we describe our prior specification for the parameters of two of these regression models.

We consider a regression setting with design matrices $\mathbf{X}$ and $\mathbf{Z}$ for the hazard rate parameters $\lambda_{awg}$ and AE probabilities $p_{ag}$ respectively. The matrices $\mathbf{X}$ and $\mathbf{Z}$ are $G \times q_{x}$ and $G \times q_{z}$ respectively, and
$\mathbf{x}_{g}^{T}$ and $\mathbf{z}_{g}^{T}$ denote the $g^{th}$ rows of $\mathbf{X}$ and $\mathbf{Z}$ respectively.
The hazard rate parameters $\lambda_{awg}$ and the adverse-event probabilities are related to $\mathbf{x}_{g}$ and $\mathbf{z}_{g}$ via
\begin{equation}
\log( \lambda_{awg} ) = \mathbf{x}_{g}^{T}\bbeta_{aw} \qquad \textrm{and} \qquad \textrm{logit}( p_{ag} ) = \mathbf{z}_{g}^{T}\bgamma_{a}, \nonumber
\end{equation}
where $\textrm{logit}(x) = \log\{x/(1-x)\}$. Here $\bbeta_{aw}$ is the $q_{x} \times 1$ vector $\bbeta_{aw} = (\beta_{aw,1}, \ldots, \beta_{aw,q_{x}})^{T}$, and $\bgamma_{aw}$ is the $q_{z} \times 1$ vector $\bgamma_{a} = (\gamma_{a,1}, \ldots, \gamma_{a,q_{z}})^{T}$. In total, this regression model involves $4q_{x} + 2q_{z}$ parameters. We describe a few possible ways of choosing $\mathbf{X}$ and $\mathbf{Z}$ below.

\subsubsection{Example 1: Saturated Model.} \label{sss:saturated}
In this model, the number of regression parameters is equal to the number of summary statistics. The design matrices $\mathbf{X}$ and $\mathbf{Z}$ are assumed to be equal to the $G \times G$ identity matrix, and each $\bbeta_{aw}$, $\bgamma_{a}$
are assumed to be $G \times 1$ vectors. This means that, for any combination of $(a,w,g)$, we have
\begin{equation}
\log(\lambda_{awg}) = \beta_{aw,g} \qquad \textrm{and} \qquad
\textrm{logit}(p_{ag}) = \gamma_{a,g}. \nonumber
\end{equation}
In this model, there is no regression structure linking the parameters $\lambda_{awg}$ or the parameters $p_{ag}$ together in such a way that more strength is borrowed between subgroups which share a greater number of patient characteristics. Rather, the $\lambda_{awg}$ and $p_{ag}$ are treated separately with no additional information used to indicate the relationships among the subgroups. 

\subsubsection{Example 2: Additive Model.} \label{sss:additive}
In this model, the values of $\lambda_{awg}$ and $p_{ag}$ are determined additively from the variables comprising subgroup $g$. This approach is analogous to the Dixon and Simon model for subgroup analysis (see e.g., \cite{dixonsimon:1991}, \cite{jones:2011}, or \cite{henderson:2016}). The number of regression parameters in this model are determined by both the number of patient variables and the number of levels within each one of these variables. 

To define the additive model, we suppose that the $G$ subgroups are formed by $p$ patient variables, and the $j^{th}$ patient variable has levels $1,\ldots,p_{j}$. If we consider modeling $\lambda_{awg}$ for subgroup $g$, it is necessary to know the levels of each of the patient variables comprising subgroup $g$. If the level of the $j^{th}$ variable in subgroup $g$ is denoted by $g(j)$, then 
\begin{eqnarray}
\log( \lambda_{awg} ) &=& \beta_{aw,1} + \sum_{j=1}^{p}\sum_{k=2}^{p_{j}} \mathbf{1}\{ g(j) = k \}\beta_{aw,Q_{j} + k}  \nonumber \\
\textrm{logit}(p_{ag}) &=& \gamma_{a,1} + \sum_{j=1}^{p}\sum_{k=2}^{p_{j}} \mathbf{1}\{ g(j) = k \}\gamma_{a,Q_{j} + k}, \label{eq:anova_model}
\end{eqnarray}
where $Q_{1} = 0$ and $Q_{j} = 1 - j + p_{1} + \ldots + p_{j-1}$ for $j > 1$. As an illustration of (\ref{eq:anova_model}), consider an example where four subgroups are formed from the combinations of the patient variables of age (young/old) and sex (male/female) and that male and young denote the first levels of the age and sex variables respectively. Suppose further that we label the four subgroups in the following manner: subgroup 1 denotes male/young (i.e., $g(1) = 1$, $g(2)=1$), subgroup 2 denotes male/old (i.e., $g(1)=1$, $g(2)=2$), subgroup 3 denotes female/young (i.e., $g(1)=2$, $g(2)=1$), and subgroup $4$ denotes female/old (i.e., $g(1) = 2$, $g(2)=2$). In this case, the $\lambda_{awg}$ would be expressed as
\begin{eqnarray}
\textrm{(Male/Young)} \qquad \log(\lambda_{aw1}) &=& \beta_{aw,1}  \nonumber \\
\textrm{(Male/Old)} \qquad \log(\lambda_{aw2}) &=& \beta_{aw,1} + \beta_{aw,2} \nonumber \\ 
\textrm{(Female/Young)} \qquad \log(\lambda_{aw3}) &=& \beta_{aw,1} + \beta_{aw,3} \nonumber \\ 
\textrm{(Female/Old)} \qquad \log(\lambda_{aw4}) &=& \beta_{aw,1} + \beta_{aw,2} + \beta_{aw,3}, \nonumber 
\end{eqnarray}
and the relation between the $p_{ag}$ and $(\gamma_{a,1},\gamma_{a,2},\gamma_{a,3})$ would be expressed similarly.


\subsubsection{A Proportional Hazards Model}
An even more parsimonious approach than the regression models discussed above is to assume instead that the hazard ratios $\lambda_{a1g}/\lambda_{a0g}$ do not vary across subgroups. Under this assumption, the parameters $(\lambda_{a0g}, p_{ag})$ would be modeled as
\begin{equation}
\log(\lambda_{a0g}) = \mathbf{x}_{g}^{T}\bbeta_{a} \qquad \textrm{and} \qquad \textrm{logit}(p_{ag}) = \mathbf{z}_{g}^{T}\bgamma_{a}, \nonumber
\end{equation}
and the hazard rates $\lambda_{a1g}$ would be obtained from the AE-free hazard rates $\lambda_{a0g}$ via $\lambda_{a1g} = \phi\lambda_{a0g}$ where $\phi$ is the constant hazard ratio. With this proportional hazards assumption, there are now $2q_{x} + 2q_{z} + 1$ regression parameters rather than the $4q_{x} + 2q_{z}$ parameters required for the regression models discussed in Sections \ref{sss:saturated} and \ref{sss:additive}. While this approach makes a quite restrictive assumption about how the AE-present and AE-free hazard rates are related,
this could be a useful approach if the proportional hazards assumption $\lambda_{a1g} = \phi\lambda_{a0g}$ 
is justifiable and is seen to be reasonable in light of model diagnostics.

\subsection{Prior Specification}
\subsubsection{Prior for Saturated Model.}
For each value of $(a,w)$,  we assume the parameters $\beta_{aw,g} = \log(\lambda_{awg})$ are drawn from a common normal distribution with mean $\mu_{aw}$ and variance $\tau_{aw}^{2}$
\begin{equation}
\beta_{aw,1}, \ldots, \beta_{aw,G}|\mu_{wa}, \tau_{aw} \sim \textrm{Normal}(\mu_{aw}, \tau_{aw}^{2}). \nonumber 
\end{equation}
The distribution of all the parameters $\beta_{aw,g}; w=0,1; a=0,1; g=1,\ldots G$ depends only on the following four vectors: $\bmu_{0} = (\mu_{00},\mu_{10})^{T}$, $\bmu_{1} = (\mu_{01}, \mu_{11})$, $\btau_{0} = (\tau_{00}, \tau_{10})^{T}$, and $\btau_{1} = (\tau_{01}, \tau_{11})^{T}$.  
The mean vector $\bmu_{a}$ for treatment arm $a$ is assumed to have the following bivariate normal distribution
\begin{equation}
\bmu_{a} \sim MVN_{2}\Bigg(\begin{bmatrix} 0 \\ 0 \end{bmatrix}, \begin{bmatrix} \sigma_{\mu,a}^{2} & \sigma_{\mu,a} \rho_{\mu,a}^{2} \\
\sigma_{\mu,a}^{2} \rho_{\mu,a} & \sigma_{\mu,a}^{2} \end{bmatrix} \Bigg), \nonumber
\end{equation}
where $\mathbf{Z} \sim MVN_{2}(\bmu, \bSigma)$ means that $\mathbf{Z}$ has a bivariate normal distribution with mean vector $\bmu$ and covariance matrix $\bSigma$. Similarly, the distribution for $\btau_{a}$ on the log scale is assumed to be given by
\begin{equation}
\log(\btau_{a}) \sim MVN_{2}\Bigg( \begin{bmatrix} \log(1/2) \\ \log(1/2) \end{bmatrix}, \begin{bmatrix} \sigma_{\tau,a}^{2} & \sigma_{\tau,a}^{2} \rho_{\tau,a} \\
\sigma_{\tau,a}^{2} \rho_{\tau,a} & \sigma_{\tau,a}^{2} \end{bmatrix} \Bigg). \nonumber 
\end{equation}
For the correlation parameters $\rho_{\mu,a}$ and $\rho_{\tau,a}$, we assume that $\rho_{\mu,a} \sim \textrm{Uniform}(-1,1)$ and $\rho_{\tau,a} \sim \textrm{Uniform}(-1,1)$ for each $a$.
Our prior distribution for the $\gamma_{a,g}$ is similar to the prior for the $\beta_{aw,g}$. Specifically, we start by assuming that
\begin{equation}
\gamma_{a,1}, \ldots, \gamma_{a,G}|\mu_{a,\gamma}, \tau_{a} \sim \textrm{Normal}(\mu_{a,\gamma}, \tau_{a,\gamma}^{2}). \nonumber
\end{equation}
We then assume that $\mu_{a,\gamma} \sim \textrm{Normal}(\log(1/2),\sigma_{\mu,a,\gamma}^{2})$ for $a=0,1$, and we assume that $\log(\tau_{a,\gamma}) \sim \textrm{Normal}(0,\sigma_{\tau,a,\gamma}^{2})$ for $a = 0,1$. 

In total, there are $8$ hyperparameters in this model. These are: $\sigma_{\mu,0}$, $\sigma_{\mu,1}$, $\sigma_{\tau,0}$, $\sigma_{\tau,1}$, $\sigma_{\mu,0,\gamma}^{2}$, $\sigma_{\mu,1,\gamma}$, $\sigma_{\tau,0,\gamma}$, and $\sigma_{\tau,1,\gamma}$. Our default choice is to set $\sigma_{\mu,0} = \sigma_{\mu,1} = \sigma_{\mu,0,\gamma} = \sigma_{\mu,1,\gamma} = 100$ and to set $\sigma_{\tau,0} = \sigma_{\tau,1} = \sigma_{\tau,0,\gamma} = \sigma_{\tau,1,\gamma} = 1$. Our justification for these default settings is discussed in more detail below.

Our prior specification implies the $\log(\lambda_{awg}), g = 1,\ldots,G$ are drawn from a common normal distribution with mean $\mu_{aw}$ and variance $\tau_{aw}^{2}$. Because it is difficult to specify a range of reasonable values for the median of $\log(\lambda_{awg}), g = 1,\ldots, G$, we assign $\mu_{aw}$ a vague prior distribution by setting $\sigma_{\mu,a} = 100$, for $a = 0, 1$. A similar justification can be made for setting $\sigma_{\mu,a,\gamma} = 100$.
Turning now to the variation in $\log(\lambda_{awg})$ we note that, for any pairs of subgroup indices $j,k$, the distribution of the hazard ratio (given $\tau_{aw}$) $\lambda_{awj}/\lambda_{awk}$ follows a log-normal distribution with parameters $0$ and $2\tau_{aw}^{2}$. Our prior distribution for $\tau_{aw}$ is based on the observation that the hazard ratio is most likely between $1/4$ and $4$. Because $P\{ \tfrac{1}{4} \leq \lambda_{awj}/\lambda_{awk} \leq 4|\tau_{aw} \} \approx 1 - 2\Phi(-1/\tau_{aw})$, this means that this approximate probability is greater than $0.68$ whenever $\tau_{aw} \leq 1$ and greater than $0.95$ whenever $\tau_{aw} \leq 1/2$. We want to choose a prior for $\tau_{aw}$ such that $P\{ \tfrac{1}{4} \leq \lambda_{awj}/\lambda_{awk} \leq 4|\tau_{aw} \}$ is fairly large for the most probable values of $\tau_{aw}$. This can be done by assuming that $\tau_{aw}$ follows a log-normal distribution with parameters $\log(1/2)$ and $1$ (i.e., setting $\sigma_{\tau,a} = 1)$. This means that $\tau_{aw}$ has a median of $1/2$ and that the probability that $\tau_{aw}$ is less than $2.6$ is approximately $0.95$. A similar justification can be made for setting   
$\sigma_{\tau,a,\gamma} = 1$.

\subsubsection{Prior for Additive Model.}
We assign vague priors to the intercept terms $\beta_{aw,1}$ and $\gamma_{a,1}$. Specifically, it is assumed that
\begin{equation}
\beta_{aw,0} \sim \textrm{Normal}(0, \sigma_{0,\beta}^{2})
\qquad \textrm{and} \qquad \gamma_{a,0} \sim \textrm{Normal}(0, \sigma_{0,\gamma}^{2}), \nonumber
\end{equation}
with $\sigma_{0,\beta}$ and $\sigma_{0,\gamma}$ set to default values of $\sigma_{0,\beta} = 100$ and $\sigma_{0, \gamma} = 100$.

The hierarchical model for the remaining regression coefficients $\beta_{aw,j}, j = 1,\ldots, \sum p_{j} - (p-1)$ and $\gamma_{a,j}, j= 2,\ldots, \sum p_{j} - (p - 1)$ is the same as that used for the regression coefficients in the saturated model. That is, we assume that
\begin{eqnarray}
\beta_{aw,1},\ldots,\beta_{aw,q_{x}} &\sim& \textrm{Normal}(\mu_{aw}, \tau_{aw}^{2}) \nonumber \\
\gamma_{a,1}, \ldots, \gamma_{a, q_{z}} &\sim& \textrm{Normal}(\mu_{a,\gamma}, \tau_{a,\gamma}^{2}), \nonumber
\end{eqnarray}
where $q_{x} = q_{z} = \sum_{j=1}^{p} p_{j} - (p-1)$. The prior distributions and default hyperparameters for $\mu_{a,\gamma}$, $\tau_{a,\gamma}$ and the four vectors $\bmu_{0} = (\mu_{00},\mu_{10})^{T}$, $\bmu_{1} = (\mu_{01}, \mu_{11})$, $\btau_{0} = (\tau_{00}, \tau_{10})^{T}$, and $\btau_{1} = (\tau_{01}, \tau_{11})^{T}$ are exactly the same as described for the saturated model. 

\subsubsection{Posterior Computation}
The Bayesian bivariate subgroup models proposed here can be easily computed using STAN software. With STAN, we have implemented both the saturated and additive models. In our analysis of the SPRINT trial, we used $4,000$ posterior draws obtained from four chains each having $1500$ iterations with the first $500$ iterations treated as burn-in. The code for implementing our models can be obtained from our website: 
\url{http://hteguru.com/index.php/bbsga/}

\section{Measures of Risk-Benefit}
\subsection{Heterogeneity in Joint Binary Outcomes}
A natural bivariate outcome of interest is survival past a specified time point $\kappa_{0}$ coupled with an indicator of whether or not an AE occurred during the follow-up period. For this bivariate outcome, there are four possible results: survival past $\kappa_{0}$/no AE occurs, survival past $\kappa_{0}$/AE occurs, failure before $\kappa_{0}$/no AE occurs, and failure before $\kappa_{0}$/AE occurs. The subgroup-specific posterior probabilities of each of these four outcomes may be obtained from the models detailed in Sections 3 and 4. The differences in these probabilities between treatment arms provide a measure by which to compare how treatment influences the probability of experiencing each of these four outcomes. If we define the function $F_{ag}(t,w)$ as
\begin{eqnarray}
F_{ag}(t,w) &=& P(T_{i} \leq t, W_{i}=w|A_{i}=a, G_{i}=g) \nonumber \\
&=& (1 - \exp(-\lambda_{awg} t))p_{ag}^{w}(1 - p_{ag})^{1-w}, \nonumber
\end{eqnarray}
the differences in these four joint probabilities may be expressed as 
\begin{eqnarray}
\theta_{g1} = S_{1g}(\kappa_{0}, 0) &-& S_{0g}(\kappa_{0},0) \nonumber \\
\theta_{g2} = S_{1g}(\kappa_{0}, 1) &-& S_{0g}(\kappa_{0}, 1) \nonumber \\
\theta_{g3} = F_{1g}(\kappa_{0}, 0) &-& F_{0g}(\kappa_{0},0) \nonumber \\
\theta_{g4} = F_{1g}(\kappa_{0}, 1) &-& F_{0g}(\kappa_{0}, 1) \nonumber 
\end{eqnarray}
If the PE is regarded as more undesirable than the adverse event, then the parameters $(\theta_{g1}, \theta_{g2}, \theta_{g3}, \theta_{g4})$ may be thought of as being ordered according to the desirability of the associated outcome. Specifically,
$\theta_{g1}$ represents the difference in the probability of achieving the most desirable outcome, namely remaining both PE and AE free until time $\kappa_{0}$. Similarly, 
$\theta_{g2}$ represents the difference in the probability of achieving the second most desirable outcome, namely remaining free from the PE until time $\kappa_{0}$ while experiencing an AE at some time before $\kappa_{0}$. The parameter $\theta_{g3}$ represents the difference in probability of experiencing the PE while not experiencing the AE, and $\theta_{g4}$ represents the difference in probability of experiencing both the primary and adverse event before time $\kappa_{0}$.
It is worth noting that $\sum_{j=1}^{4} \theta_{gj} = 0$ for each $g$.
That is, if the treatment increases the probability of one outcome relative to control, it must decrease the probability of another outcome (or group of outcomes) by a corresponding amount.

\begin{figure}
    \begin{center}
             \includegraphics[width=14cm, height=14cm]{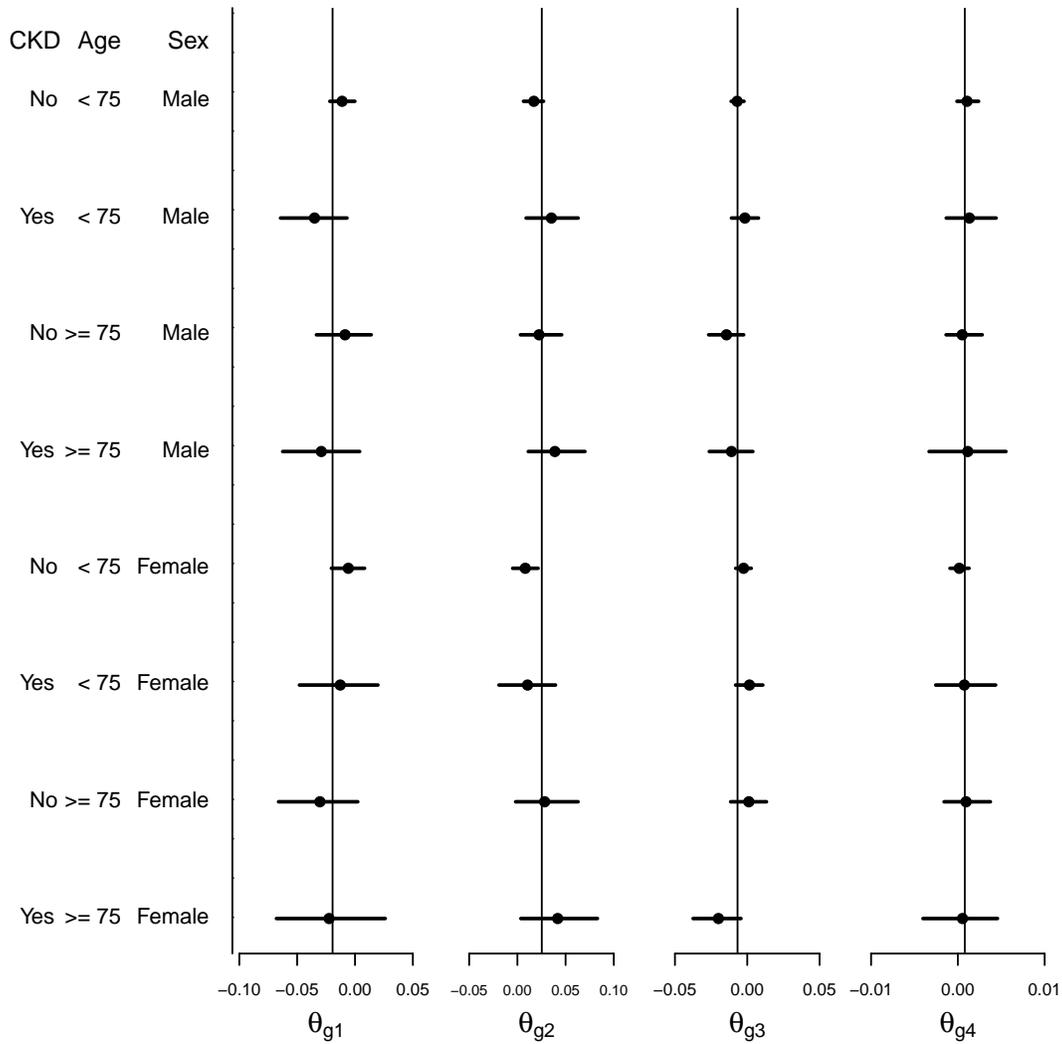}
     \caption{SPRINT trial with eight subgroups defined by chronic kidney disease (CKD), age, and sex. Posterior means and associated credible intervals using the saturated model are plotted for $\theta_{g1}$, $\theta_{g2}$, $\theta_{g3}$, $\theta_{g4}$. Note that the range of the x-axis can differ across $\theta_{g1}$, $\theta_{g2}$, $\theta_{g3}$, $\theta_{g4}$. For each $j$, the vertical line is placed at $\hat{\theta}_{j,ov}$, the average value of the posterior means $\hat{\theta}_{gj}$ across subgroups.}
\label{fig:four_forestplot}
     \end{center}
\end{figure}

Using the SPRINT trial data with the $8$ subgroups defined by chronic kidney disease status, age, and sex, Figure \ref{fig:four_forestplot} plots the posterior means of $(\theta_{g1}, \theta_{g2}, \theta_{g3}, \theta_{g4})$. The posterior quantities shown in Figure \ref{fig:four_forestplot} were computed using the saturated model. For each of the forest plots in Figure \ref{fig:four_forestplot}, the solid vertical line represents the average value of the posterior means of $\theta_{gj}$ across subgroups, i.e., $\tfrac{1}{G}\sum_{g} \hat{\theta}_{gj}$ where $\hat{\theta}_{gj}$ is the posterior mean of $\theta_{gj}$. The quantity $\hat{\theta}_{j, ov} = \tfrac{1}{G}\sum_{g=1}^{G}\hat{\theta}_{gj}$ can be thought of as an estimate of the overall treatment effect for PE-AE category $j$. Looking at Figure \ref{fig:four_forestplot}, the estimated overall treatment effect is negative for PE-AE categories 1 and 3 and is positive for categories 2 and 4 though the magnitude of the estimates for categories 3 and 4 is very small. Specifically, the values of $\hat{\theta}_{j,ov}$ were $\hat{\theta}_{1,ov} = -0.019$, $\hat{\theta}_{2,ov} = 0.025$, $\hat{\theta}_{3,ov} = -0.007$, and $\hat{\theta}_{4,ov} = 0.001$.
These values suggest the intensive treatment increases the probability of 3-year PE-free survival while experiencing a treatment-related SAE, and it reduces the probability of experiencing a PE without an SAE.
However, the intensive treatment also appears to reduce the probability of remaining both PE and SAE free for 3 years by an a similar amount to the increase in the probability of outcome $2$. Comparing the estimates $\hat{\theta}_{1,ov}$, $\hat{\theta}_{2,ov}$, and $\hat{\theta}_{3,ov}$ leads to a somewhat ambiguous picture about the effect of the intensive treatment on improving patient outcomes. A more complete evaluation of the risk/benefit effect of the intensive treatment would involve consideration of patients' ``baseline bivariate risk''. 
Specifically, the baseline bivariate risk refers to the probability of each PE-AE combination occurring under the control arm. Such baseline information would be useful in assessing the meaningfulness of any change $\theta_{gj}$ that the treatment induces on the probability of outcome $j$.

Turning now to the variation in $\theta_{gj}$ across subgroups, two subgroups from Figure \ref{fig:four_forestplot} seem to stand out. The more prominent of these is the no-CKD/Age $< 75$/Female subgroup. In this subgroup, the $95\%$ credible interval for $\theta_{g2}$ does not cover the overall treatment effect estimate $\hat{\theta}_{2,ov}$, and the credible interval for $\theta_{g1}$ nearly does not cover $\hat{\theta}_{1,ov}$. This suggests the effect of the intensive treatment on altering a patients' PE-AE outcome profile is somewhat different in the no-CKD/Age $<75$/Female subgroup than in the broader population represented by the trial. Specifically, compared to other groups in the trial, the intensive treatment for those in the no-CKD/Age $<75$/Female subgroup had almost no effect on any of the PE-AE probabilities as the estimates $\hat{\theta}_{gj}$ for this subgroup were much closer to zero when compared with their overall counterparts $\hat{\theta}_{j,ov}$.
Another subgroup worth highlighting is the no-CKD/Age $< 75$/Male subgroup. Though not as extreme as the no-CKD/Age $< 75$/Female subgroup, the estimates of $\theta_{g1}$ and $\theta_{g2}$ are modestly closer to zero than the average values of these parameters, and while the credible intervals for these parameters still covered $\hat{\theta}_{1,ov}$ and $\hat{\theta}_{2,ov}$, most of the mass of these credible intervals were to the right and left of $\hat{\theta}_{1,ov}$ and $\hat{\theta}_{2,ov}$ respectively.

\begin{figure}
    \begin{center}
             \includegraphics[width=14cm, height=14cm]{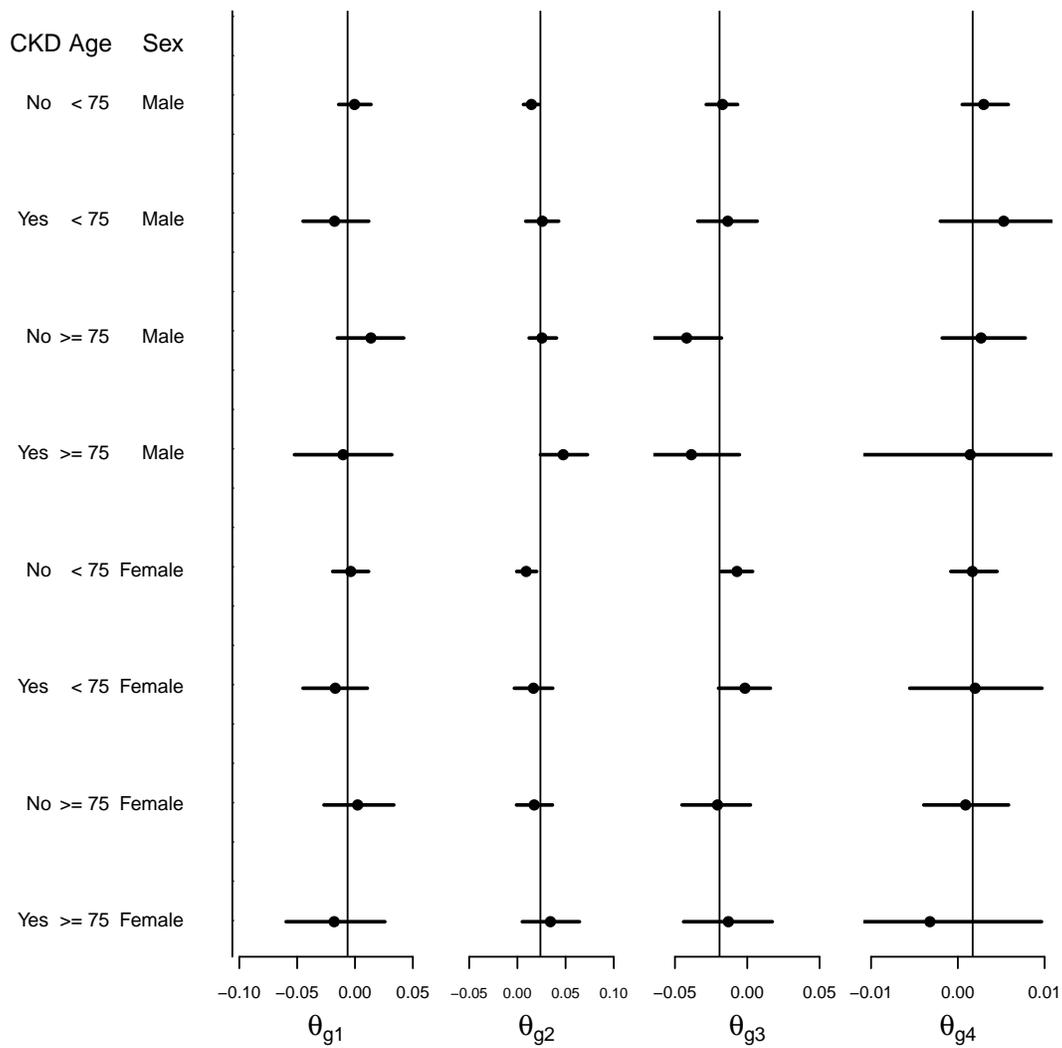}
     \caption{SPRINT trial with eight subgroups defined by chronic kidney disease (CKD), age, and sex. Posterior means and associated credible intervals using the additive model are plotted for $\theta_{g1}$, $\theta_{g2}$, $\theta_{g3}$, $\theta_{g4}$. For each $j$, the vertical line is placed at $\hat{\theta}_{j,ov}$, the average value of the posterior means across subgroups.}
\label{fig:four_forestplot_additive}
     \end{center}
\end{figure}

Figure \ref{fig:four_forestplot_additive} displays posterior estimates of $(\theta_{g1}, \theta_{g2}, \theta_{g3}, \theta_{g4})$ from the additive model. One notable difference between the additive and the saturated model is in the overall estimates $\hat{\theta}_{1,ov}$ and $\hat{\theta}_{3,ov}$. Specifically, the estimate $\hat{\theta}_{1,ov}$ was much closer to zero in the additive model than in the saturated model. 
As with the saturated model, both the no-CKD/Age $<75$/Female and no-CKD/Age $< 75$/Male subgroups stand out in the additive model due to lower magnitudes of $\theta_{g2}$. The estimated values of $\theta_{g1}$ for the  no-CKD/Age $<75$/Female subgroup is actually quite similar to those from the saturated model. However, due to differences between the saturated and additive models in the overall estimate $\hat{\theta}_{1,ov}$, the value of $\theta_{g1}$ for this subgroup does not appear different than the estimated overall value of this parameter.

\subsection{Comparing Utilities across Subgroups} \label{utility}
While forest plots of $\theta_{g1},\ldots,\theta_{g4}$ can be useful for examining variability in the effect of treatment on altering patients' risk-benefit profiles, interpreting such graphs can be somewhat challenging. Converting the bivariate outcomes into a composite score by weighting each outcome allows one to report heterogeneity with respect to a univariate score. 
Weighting methods for assessing risk-benefit have been described by a number of researchers including, for example, \cite{chuang:1994}.
We consider here two approaches for weighting joint patient outcomes.

A direct way of combining $\theta_{g1}, \ldots, \theta_{g4}$ to produce a univariate score would be to compute 
\begin{equation} \label{ttg}
\tilde{\theta}_{g} = b_{1}\theta_{g1} + b_{2}\theta_{g2} + b_{3}\theta_{g3} + b_{4}\theta_{g4}
\end{equation}
for weights $b_{1}, \ldots, b_{4}$. Assuming that a patient receives a utility of $b_{j}$ when experiencing outcome $j$, $\tilde{\theta}_{g}$ represents the difference in expected utility between treatments for those in subgroup $g$. In other words, if one defines a composite outcome where a patient is assigned a composite of score of $b_{j}$ according to , then $\tilde{\theta}_{g}$ represents the subgroup-specific expected difference in this composite outcome.

Rather than weight joint binary outcomes as is done in computing $\tilde{\theta}_{g}$, an alternative is to incorporate the survival time directly into the composite outcome. We define such a composite score $H_{i}$ which, for the $i^{th}$ patient, we define as
\begin{equation}
H_{i} = b_{1}W_{i}\min\{ T_{i}, \tau \} + b_{2}(1 - W_{i})\min\{ T_{i}, \tau\}, 
\end{equation}
where $\tau$ represents a truncation point that represents a follow-up period of interest. The composite measure $H_{i}$ may be viewed as an outcome whose expectation is a type of weighted restricted mean survival time (RMST) (see e.g., \cite{royston:2013}). We use $\min\{ T_{i}, \tau\}$ rather than $T_{i}$ to reduce the impact of model extrapolation beyond the follow-up period of interest $(0, \tau)$.
With the measure $H_{i}$, a patient receives a weight of $b_{2}$ for each time unit of PE-free survival assuming he/she did not experience an AE over the time interval $(0, \min\{T_{i}, \tau\})$. Likewise, a patient receives a weight of $b_{1}$ for each unit of PE-free survival time assuming he/she did experienced an adverse event at some point in the time interval $(0, \min\{T_{i}, \tau\})$.   
The composite outcome $H_{i}$ bears some resemblance to the Q-TWist measure described in \cite{Gelber:1989} though $H_{i}$ does not explicitly distinguish between time with toxicity and time without toxicity. 

\begin{figure}
    \begin{center}
             \includegraphics[width=14cm, height=14cm]{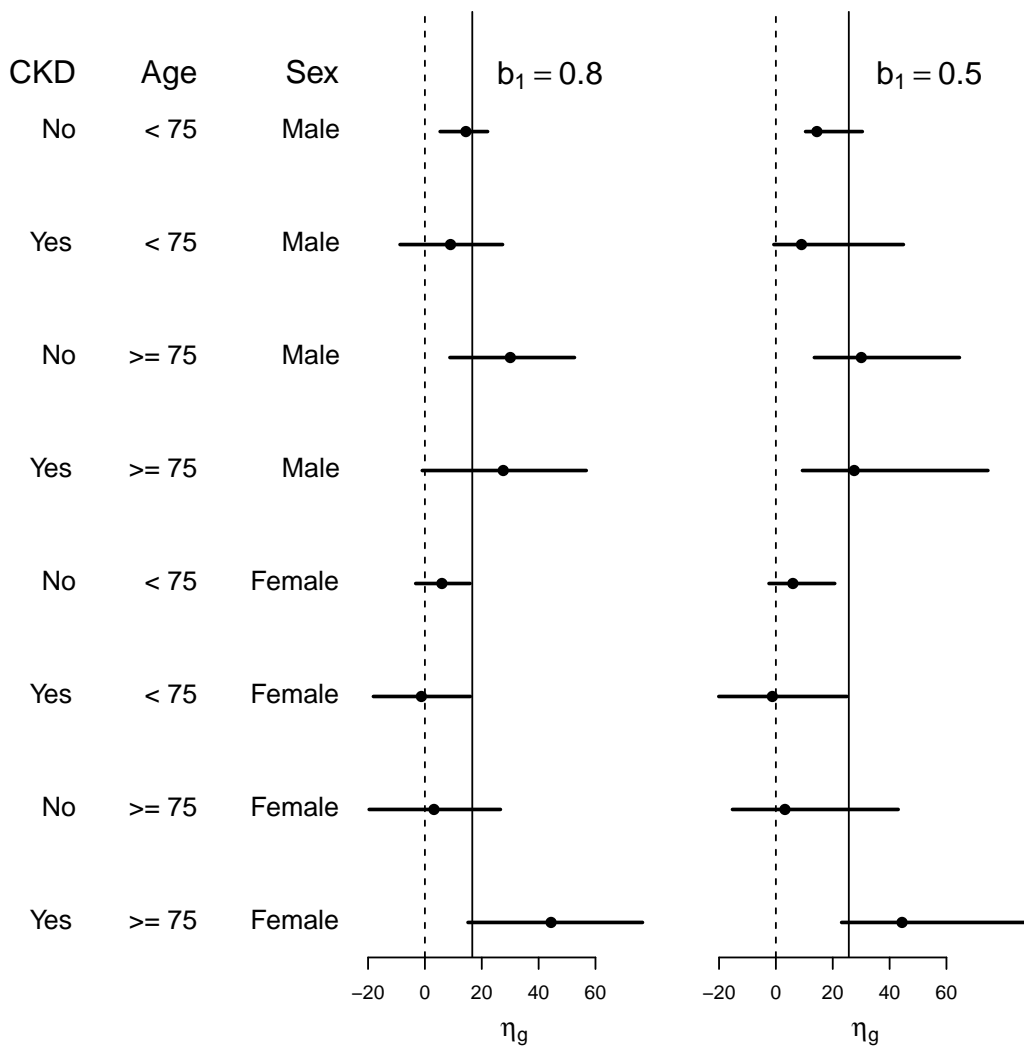}
     \caption{SPRINT trial with eight subgroups defined by chronic kidney disease (CKD), age, and sex. Utility analysis using the saturated model and composite outcome measure $H_{i}$. Posterior means and credible intervals for $\eta_{g}$ are shown for each subgroup. The solid vertical line is placed at the average value of the posterior means of $\eta_{g}$ across subgroups.}
\label{fig:utility_saturated}
     \end{center}
\end{figure}

The expectation of the composite outcome $H_{i}$ conditional on subgroup and treatment arm information is given by
\begin{eqnarray}
E[H_{i}|A_{i}=a, G_{i}=g] &=& \frac{b_{1}p_{ag}(1 - e^{-\tau\lambda_{a1g}})}{\lambda_{a1g}} + \frac{b_{2}(1 - p_{ag})(1 - e^{-\tau\lambda_{a0g}})}{\lambda_{a0g}}. \nonumber
\end{eqnarray}
Subgroup-specific treatment effects $\eta_{g}$ relative to this measure may then be defined as the difference in the expectation of $H_{i}$ in the treatment arm vs. the expectation of $H_{i}$ in the control arm
\begin{equation} \label{etag}
\eta_{g} = E[H_{i}|A_{i}=1, G_{i}=g] - E[H_{i}|A_{i}=0, G_{i}=g]. 
\end{equation}

Figure \ref{fig:utility_saturated} shows estimates of the treatment effects $\eta_{g}$ using the saturated model and two choices of weights $(b_{1}, b_{2})$.
For each choice of weights, we set $b_{2} = 1$, and we set $b_{1}$ to $0.8$ and $0.5$ for the left-hand and right-hand plots respectively. When $b_{2} = 1$, $b_{2}$ may be interpreted as the proportion of value that one receives from a unit of extra life when an AE is known to occur as compared to the value of an extra unit of life when no AE occurs. Hence, a value of $b_{1}$ very close to one implies that AE-free survival and AE-occurring survival are valued similarly while values of $b_{1}$ much lower than one imply that AE-free survival is valued much more highly than AE-occurring survival.  

It is interesting to note that the younger subgroups (i.e., age $< 75$) tend to consistently receive less benefit treatment for both choices of weights. Indeed, all of the younger subgroups have posterior means less than the average value of the posterior means $\hat{\eta}_{g}$.
In Figure \ref{fig:utility_saturated}, two subgroups stand out in terms of being different than the average value of $\eta_{g}$ across subgroups. These are the no-CKD/Age $<75$/Female and yes-CKD/Age $< 75$/Female subgroups. Both of these subgroups do not appear to benefit when using either choice of weights to compute $H_{i}$. This is not that surprising for the no-CKD/Age $< 75$/Female subgroup as the four-outcome subgroup analysis from Figure \ref{fig:four_forestplot} suggested the intensive treatment had very little impact in changing the probabilities of any of the four outcomes. Similarly, the point estimates from Figure \ref{fig:four_forestplot} also suggested that the treatment had little effect on the yes-CKD/Age $< 75$/Female subgroup though the posterior uncertainty for these estimates was much greater than for the no-CKD/Age $< 75$/Female subgroup. Interestingly, the no-CKD/Age $<75$/Male subgroup did not show clear evidence of having a worse treatment effect $\eta_{g}$ than the average from the other subgroups.

\subsection{Heterogeneity in the Probability of Outcome Improvement}
While using weights to create univariate composite measures can 
be useful, it is often difficult to justify a particular choice of weights in an analysis. An alternative to weighting is to order the outcomes according to their desirability and to use the probability of outcome improvement as the treatment effect to be measured. An advantage of approaches which rely on ordering outcomes (see e.g., \cite{claggett:2015} or \cite{follmann:2002}) is that they do not require that one assign numerical values to each outcome but only require that one be able to order outcomes according to their desirability. 

We consider here an ordering-based definition of treatment effect that compares the outcomes of two patients randomly drawn from each treatment arm. Specifically, we consider two treatment-discordant patients $i$ and $j$ (say $A_{i} = 1$ and $A_{j} = 0$) which are chosen randomly from subgroup $g$. If we could observe the failure times $T_{i}$ and $T_{j}$ along with $W_{i}$ and $W_{j}$, it would be possible to determine which patient had the superior outcome. That is, we could construct a function $O(\cdot, \cdot|\cdot,\cdot)$ such that $O(T_{i}, W_{i}|T_{j}, W_{j}) = 1$ when outcome $(T_{i}, W_{i})$ is superior to $(T_{j}, W_{j})$ and $O(T_{i}, W_{i}|T_{j}, W_{j}) = 0$ otherwise. While there are many ways of choosing $O(T_{i}, W_{i}|T_{j}, W_{j})$, Table \ref{tab:outcome_ordering} describes our approach for ordering pairs of bivariate outcomes. Here, we rely on an indifference parameter $\delta > 0$ which represents the additional gain in survival that would be needed to compensate for suffering from the AE. 
For example, suppose that $W_{i} = 1$ and $W_{j} = 0$ and that the indifference parameter is set to $\delta = 0.5$. In this case, patient $i$ would need an at least $50\%$ longer survival time than patient $j$ in order for the outcome of patient $i$ to be preferable to the outcome of patient $j$.

\begin{table}[ht]
\centering
\begin{tabular}{lrc} \toprule
 \multicolumn{2}{c}{Outcomes} & \multicolumn{1}{c}{Preferred Treatment} \\
\midrule 
$T_{i} > T_{j}(1 + \delta)$ & $W_{i} = 1,  W_{j} = 0$ & $A = 1$ \\
$T_{i} \leq T_{j}(1 + \delta)$ & $W_{i} = 1,  W_{j} = 0$ & $A = 0$ \\
$T_{j} > T_{i}(1 + \delta)$ & $W_{i} = 0,  W_{j} = 1$ & $A = 0$ \\
$T_{j} \leq T_{i}(1 + \delta)$ & $W_{i} = 0,  W_{j} = 1$ & $A = 1$ \\
$T_{i} > T_{j}$ & $W_{i} = 1, W_{j} = 1$ & $A = 1$ \\
$T_{i} > T_{j}$ & $W_{i} = 0, W_{j} = 0$ & $A = 1$ \\
$T_{i} \leq T_{j}$ & $W_{i} = 1,  W_{j} = 1$ & $A = 0$ \\
$T_{i} \leq T_{j}$ & $W_{i} = 0, W_{j} = 0$ & $A = 0$ \\
\bottomrule
\end{tabular}
\caption{Our approach for comparing two joint outcomes $(T_{i}, W_{i})$ and $(T_{j}, W_{j})$ assuming that $A_{i} = 1$ and $A_{j} = 0$.}
\label{tab:outcome_ordering}
\end{table}

Using the outcome orderings described in Table \ref{tab:outcome_ordering}, we define the subgroup-specific treatment effects $\phi_{g}$ as
\begin{equation} \label{phig}
\phi_{g} = 2P\big\{ O(T_{i}, W_{i}|T_{j}, W_{j})= 1 \mid A_{i} = 1, A_{j} = 0, G_{i}=g, G_{j} = g \big\} - 1. 
\end{equation}
The parameter $\phi_{g}$ represents the probability that patient $i$ has a superior outcome to patient $j$ minus the probability that patient $j$ has a superior outcome to patient $i$ when assuming both patients $i$ and $j$ belong to subgroup $g$. Hence, $\phi_{g} = 1$ when patients in subgroup $g$ are sure to benefit from treatment, and $\phi_{g} = -1$ when patients in subgroup $g$ are certain to be harmed by treatment. 
Note that $\phi_{g}$ is a function of the subgroup-specific parameters $\lambda_{awg}$ and $p_{ag}$ and thus one may easily obtain draws from the posterior distribution of $\phi_{g}$ by transforming samples from the posterior distribution of $(\lambda_{awg}, p_{ag})$. More specifically, $\phi_{g}$ is related to the model parameters via
\begin{equation}
\phi_{g} = 2\Big[ \frac{\lambda_{00g}p_{1g}(1 - p_{0g})}{\lambda_{11g}(1 + \delta) + \lambda_{00g}} + \frac{\lambda_{01g}(1 - p_{1g})p_{0g}}{\lambda_{10g}/(1 + \delta) + \lambda_{01g}} + \frac{\lambda_{00g}(1 -p_{1g})(1 - p_{0g})}{\lambda_{10g} + \lambda_{00g}} + \frac{\lambda_{01g}p_{1g}p_{0g}}{\lambda_{11g} + \lambda_{01g}} \Big] - 1. \nonumber
\end{equation}

\begin{figure}
    \begin{center}
             \includegraphics[width=14cm, height=14cm]{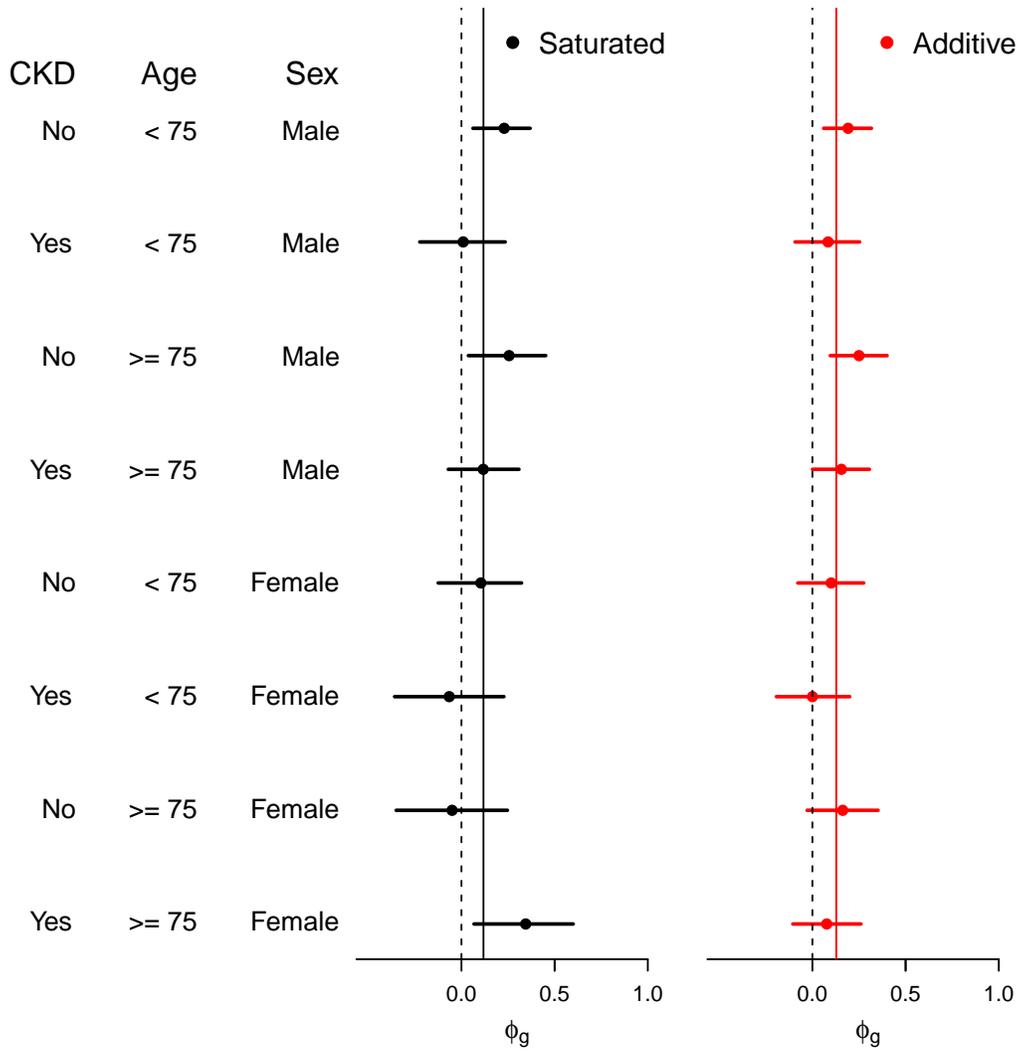}
     \caption{Sprint Trial: Posterior means of treatment effects $\phi_{g}$ with $\delta = 0.2$ (see Table \ref{tab:outcome_ordering}). Posterior means and credible intervals for $\phi_{g}$ are shown for both the saturated and additive models. The solid vertical lines are placed at the overall treatment effect estimates i.e, $\tfrac{1}{G}\sum_{g=1}^{G} \hat{\phi}_{g}$, where $\hat{\phi}_{g}$ is the posterior mean of $\phi_{g}$.}
\label{fig:ranking_forest}
     \end{center}
\end{figure}

Figure \ref{fig:ranking_forest} plots the posterior means of $\phi_{g}$ for each subgroup using both the saturated and additive models. For this analysis, we set $\delta$ to $\delta = 0.2$. This means that an individual receiving the intensive treatment would need at least $1.2$ times as much PE-free survival as under the standard treatment in order to be ``compensated'' for the fact that he/she would experience a treatment-related SAE under the intensive treatment while remaining SAE-free under the standard treatment arm. In Figure \ref{fig:ranking_forest}, the ``overall'' treatment effects are represented by the solid vertical lines.
These are computed as $\tfrac{1}{G}\sum_{g=1}^{G} \hat{\phi}_{g}$ where $\hat{\phi}_{g}$ is the posterior mean of $\phi_{g}$. The overall treatment effect estimates were $0.12$ and $0.13$ for the saturated and additive models respectively. 

As shown in Figure \ref{fig:ranking_forest}, there is less clear variability across subgroup when using the treatment effect $\phi_{g}$. Specifically, all the credible intervals for both the saturated and additive models cover the average value of the posterior means $\hat{\phi}_{g}$. It is interesting to note that, on this scale, the estimated treatment effect for the no-CKD/Age $< 75$/Female subgroup is now very close to the estimated overall treatment effect which seems to stand in contrast to the results in Sections 5.1 and 5.3. While the width of the credible intervals in both the saturated and additive models prevents us from making any strong conclusions about this subgroup, it is worthwhile to note how the prominence of a treatment-covariate interaction can change when the treatment effect scale is changed. The differences between the saturated and additive models for the no-CKD/Age $\geq 75$/Female and yes-CKD/Age $\geq 75$/Female subgroups is also noteworthy. In addition to much greater shrinkage with the additive model, the difference between the estimates of $\phi_{g}$ in these $2$ subgroups changes sign between the saturated and additive models. This is due to the more structured borrowing of information among subgroups that is enabled by the additive model. Specifically, much information is shared among subgroups carrying the same CKD status with the no-CKD subgroups being consistently shrunken towards a larger value than the yes-CKD subgroups.

\section{Model Checking and Diagnostics}
Our approach assumes that, conditional on an AE indicator, survival follows an exponential distribution within each subgroup/treatment arm combination.
We feel that such a simple parametric approach is needed due to the potentially large number of ``cells'' that arise from the number of subgroup/treatment arm/AE combinations where the sparse amount of within-cell data precludes more ambitious modeling. Despite this, our model treats patient survival as arising from a type of mixture-of-exponentials distribution and such a mixture can often provide a good fit to the observed overall survival patterns.

\begin{figure}
    \begin{center}
             \includegraphics[width=18cm, height=12cm]{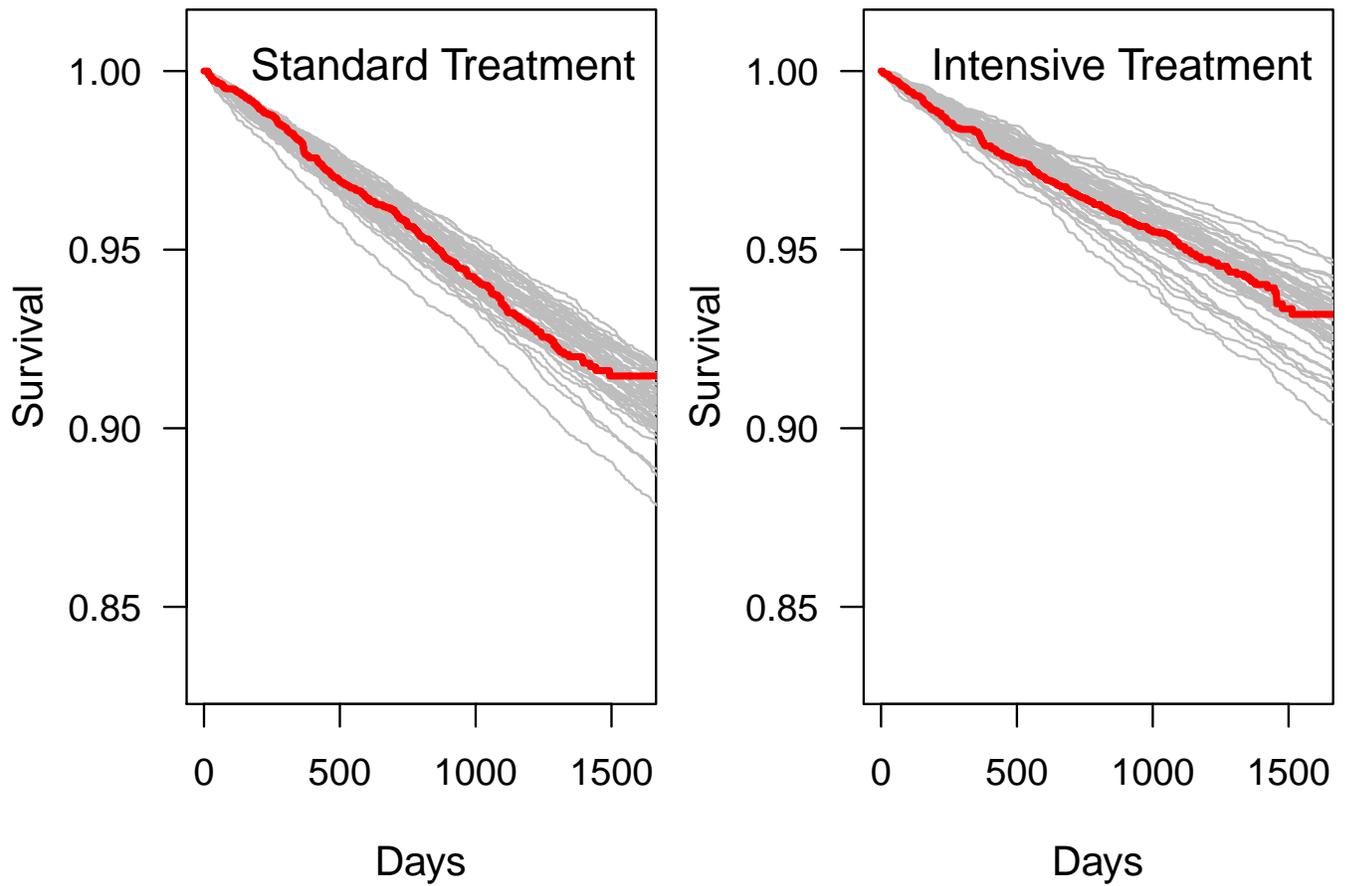}
     \caption{SPRINT trial: Draws from the posterior predictive distribution for the saturated model. Kaplan-Meier estimates for the primary outcomes from the observed data are plotted in solid red. Kaplan-Meier estimates from fifty draws from the posterior predictive distribution are plotted in grey.}
 \label{fig:pp_survival_curves}
\end{center}
\end{figure}

Posterior predictive checks (\cite{meng:1994}) are a useful tool for assessing the goodness-of-fit of a Bayesian model. Such checks are performed by first simulating outcomes from the posterior predictive distribution and computing a test statistic (or a collection of test statistics) of interest for each simulation replication. The distribution of the computed test statistics across posterior predictive draws is then compared with the observed value of this test statistic. An observed value of the test statistic that seems ``typical'' with respect to the posterior predictive distribution of this test statistic is an indication that the model provides a good fit or, at least, is not clearly deficient in some way. Figure \ref{fig:pp_survival_curves} shows treatment-specific estimated survival curves from $50$ draws from the posterior predictive distribution. Here, the test statistics of interest are the treatment-specific Kaplan-Meier estimates. As shown in Figure \ref{fig:pp_survival_curves}, the Kaplan-Meier estimates for the observed data seem typical when compared with the collection of Kaplan-Meier estimates from the $50$ posterior predictive draws. Indeed, for both treatment arms, the observed Kaplan-Meier estimates are roughly centered among the posterior predictive survival curves, and the shapes of the posterior predictive survival curves largely resemble the linear shape of the observed Kaplan-Meier estimates.

\begin{figure}
    \begin{center}
             \includegraphics[width=16cm, height=14cm]{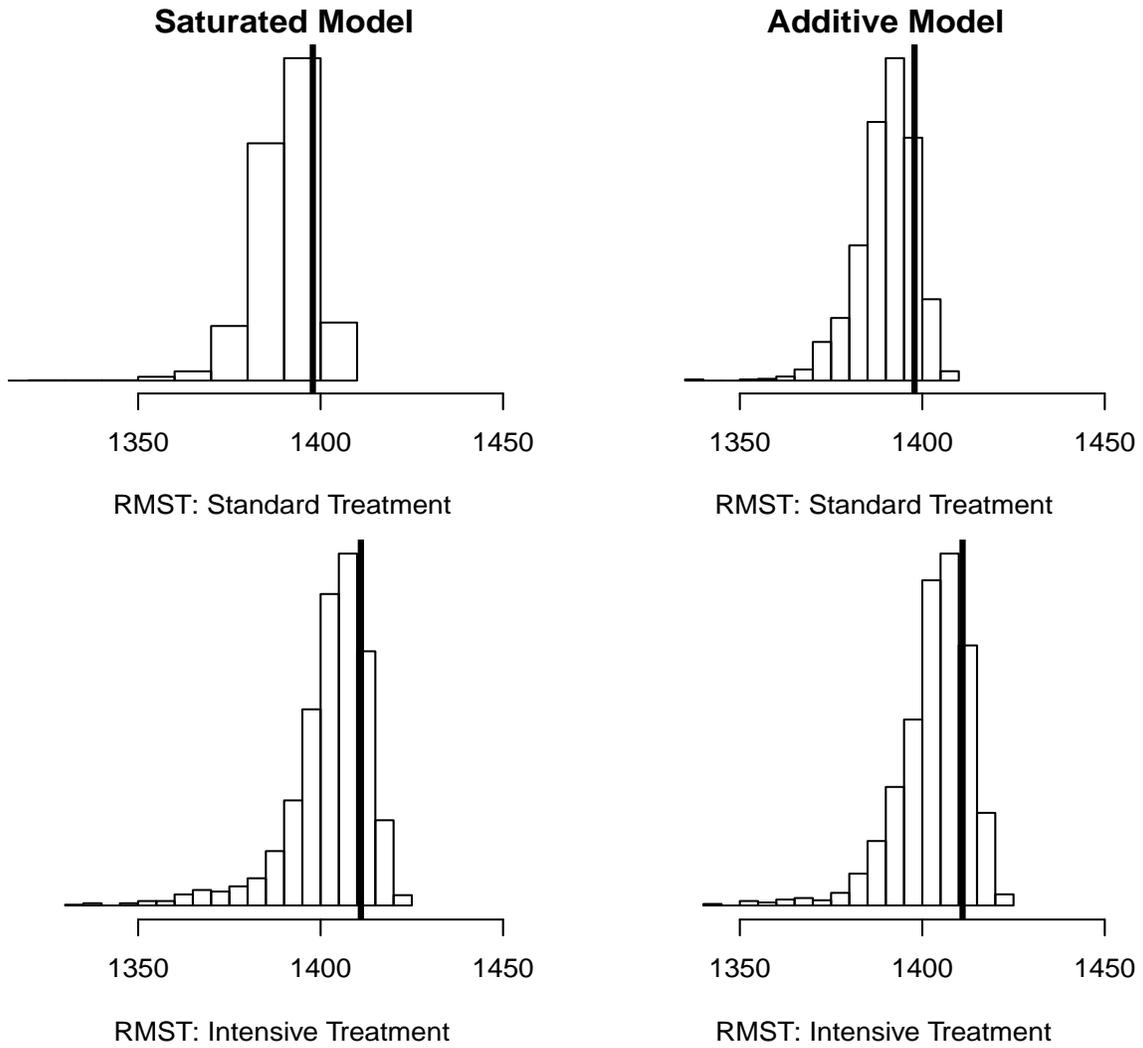}
     \caption{SPRINT trial: Posterior predictive checks. Samples from the posterior predictive distribution of the RMST (by treatment arm) using both the saturated and additive models. Solid vertical lines are placed at the observed values of the RMST.}
\label{fig:pp_rmst}
     \end{center}
\end{figure}

Beyond graphical displays of survival, one can use univariate test statistics and posterior predictive p-values as a way of more formally assessing goodness-of-fit. Posterior predictive p-values represent the probability that hypothetical replications of the test statistic from the
posterior predictive distribution is more extreme than the observed value of the test statistic. It is often recommended to try test statistics which are not directly modeled by the probability model used (\cite{gelman:2014}).
Figure \ref{fig:pp_rmst} displays the posterior predictive distributions of the restricted mean survival time (RMST) using both the saturated and additive models. For the saturated model, the two-sided posterior predictive p-values are $0.30$ and $0.38$ for the standard and intensive treatment arms respectively. For the additive model, the two-sided posterior predictive p-values are $0.28$ and $0.40$ for the standard and intensive treatment arms respectively. These relatively large posterior predictive p-values do not indicate that either model has a serious flaw.

\begin{table}[ht]
\centering
\begin{tabular}{lrc} \toprule
 \multicolumn{1}{c}{Predictive Criterion} & \multicolumn{1}{c}{Additive Model} & \multicolumn{1}{c}{Saturated Model} \\
\cmidrule(r){2-2} \cmidrule(r){3-3}
\midrule 
DIC & $12853.3$ & $12863.4$ \\
WAIC & $12861.9$ & $12865.2$ \\
\bottomrule
\end{tabular}
\label{tab:IC}
\caption{Comparisons of the saturated and additive models using different information criteria.}
\end{table}

Information criteria can be used to compare models such as the saturated and additive models that have differing number of parameters. The use of information criteria can be helpful whenever model checking procedures do not clearly show that one of the models is inadequate or when there is no strong a priori reason for favoring one model over the other. Two well-known information criteria that can be computed for Bayesian hierarchical models are the deviance information criterion (DIC) (\cite{spiegelhalter:2002}) and the widely-applicable information criterion (WAIC) (\cite{Watanabe:2010}). Both of these criterion are found by adding a penalty to negative two times the log-likelihood function evaluated at the posterior mean of the model parameters. The penalty is determined by the model degrees of freedom which are computed differently by DIC and WAIC. Lower values of both the DIC and WAIC imply better model performance.

As shown in Table \ref{tab:IC}, the additive model had a DIC which was roughly $10$ less than the DIC of the saturated model. Differences in the DIC of less that $3-5$ are often not considered meaningful (see e.g., \cite{Louis:09}) so this difference of $10$ provides moderate support in favor of the additive model. Moreover, the additive model may be especially preferred if a parsimonious model is desired. In contrast to the DIC, the saturated and additive models are much closer when using WAIC as the comparison measure. Nevertheless, a difference of roughly $4$ in WAIC value still provides mild support in favor of the additive model. Thus, for the SPRINT data, there does not seem to be a good justification for using the saturated model as the log-likelihood (evaluated at the posterior mean of the parameters) for both models is roughly the same while the saturated model requires considerably fewer model parameters.


\section{Discussion}
In this article, we have described a Bayesian approach for performing subgroup analysis with respect to both a primary and safety endpoint and have detailed its use in generating relevant within-subgroup risk-benefit measures. A key feature of our approach is that the benefit-safety endpoints are modeled jointly rather than used to perform separate subgroup analyses of benefit and safety. 

Summary assessment of risks and benefits, i.e. combining the multiple risks and benefit outcomes into a single utility measure, is a challenging problem. In section \ref{utility} we have provided a few reasonable options, Eqs. \ref{ttg}, \ref{etag}, \ref{phig}. Other summary measures are certainly possible.  It should be remarked that both the primary and safety endpoints used in the SPRINT trial are themselves composite endpoints. Hence, it may be worth decomposing these composite measures further for a more thorough assessment of benefit and risk (e.g., death and other cardiac events are treated the same).

\bibliographystyle{agsm}
\bibliography{bvsubgroup_refs}

\end{document}